\documentclass[aps,showpacs,superscriptaddress,nofootinbib,preprint]{revtex4}
\usepackage{graphicx}% Include figure files
\usepackage{amssymb}
\usepackage{amsmath}
\begin{document}
\title{Atmospheric neutrino flux at INO, South Pole and Pyh\"asalmi}

\author{M. Sajjad \surname{Athar}}
\email{sajathar@gmail.com}
\affiliation{Department of Physics, Aligarh Muslim University, Aligarh-202002, India}

\author{M. \surname{Honda}}
\email{mhonda@icrr.u-tokyo.ac.jp}
\affiliation{Institute for Cosmic Ray Research, the University of Tokyo, 5-1-5 Kashiwa-no-ha, Kashiwa, Chiba 277-8582, Japan}

\author{T. \surname{Kajita}}
\email{kajita@icrr.u-tokyo.ac.jp}
\affiliation{Institute for Cosmic Ray Research, and Kavli Institute for the Physics and the Mathematics of the Universe, the University of Tokyo,
5-1-5 Kashiwa-no-ha, Kashiwa, Chiba 277-8582, Japan}

\author{K. \surname{Kasahara}}
\email{kasahara@icrc.u-tokyo.ac.jp}
\affiliation{Research Institute for Science and Engineering, Waseda University, 3-4-1 Okubo Shinjuku-ku, Tokyo, 169-8555, Japan}

\author{S. \surname{Midorikawa}}
\email{midori@aomori-u.ac.jp}
\affiliation{Faculty of Software and Information Technology, Aomori University, Aomori, 030-0943 Japan}
\begin{abstract}
We present the calculation of the atmospheric neutrino fluxes 
for the neutrino experiments proposed at INO, South Pole and Pyh\"asalmi. 
Neutrino fluxes have been obtained using ATMNC, a simulation code for
cosmic ray in the atmosphere.
Even using the same primary flux model and the interaction model, 
the calculated atmospheric neutrino fluxes are different for 
the different sites due to the geomagnetic field.
The prediction of these fluxes in the present paper would be quite 
useful in the experimental analysis.
\end{abstract}
\maketitle  

\section{INTRODUCTION}\label{Intro}
The study of neutrino physics with atmospheric neutrinos 
has a long history with the first observations of muons produced 
by $\nu_\mu$ in deep underground laboratories of 
Kolar Gold Field(KGF) mines in India~\cite{kgf} and 
East Rand Propietary Mines (ERPM) in South Africa~\cite{erpm}. 
It was the Kamiokande~\cite{kamioka}, IMB~\cite{imb} and some 
other atmospheric neutrino experiments~\cite{others} which gave a clear evidence of a 
deficit in the atmospheric muon neutrino flux. With the Super-Kamiokande~\cite{sk} experiment, 
it is now well established that atmospheric neutrinos do oscillate. 
Both theoretically and 
experimentally lots of activities 
are going on to precisely determine neutrino oscillation parameters and 
experiments are planned with atmospheric as well as with the accelerator, 
reactor and solar neutrinos.

India is going to establish a neutrino observatory at the Theni district 
of Tamilnadu~\cite{ino}. 
This India-based Neutrino Observatory (INO) is proposed to be installed 
at $9^{o}96^{'}7^{''}$N, $77^{o}26^{'}7^{''}$E, 
and their primary aim is to study atmospheric neutrinos in a 1300 meters deep cave. 
The detector that is planned to be used is a magnetized iron calorimeters
(ICAL) of 50kT. 
At the South Pole, PINGU(Precision IceCube Next Generation Upgrade) 
is proposed as the upgrade version of IceCube (-$90^{o},0^{o}$)~\cite{pingu}
to observe atmospheric neutrino at lower energies by increasing the 
optical modules many times of 
the present setup, under the surface of the ice. 
Another new generation underground neutrino experiment facility
called LAGUNA(Large Apparatus studying Grand Unification and Neutrino Astrophysics) is proposed at Pyhasalmi Mine~\cite{pyh} in Finland, which is
 among the deepest mines 
in Europe at $63^{o}39{'}$N, $26^{o}02{'}$E.

In this paper, 
we have calculated atmospheric neutrino fluxes for these three sites 
in a 3D scheme using 
ATMNC (ATmospheric Muon Neutrino Calculation)
code\cite{honda1,honda2,honda3} for the cosmic ray propagation in 
atmosphere with JAM, 
which is used in PHITS 
(Particle and Heavy-Ion Transport code System)~\cite{phits},
in the hadronic interaction at lower energies ($<$ 32~GeV). 
The JAM interaction model agrees with the HARP experiment~\cite{harp} 
a little better than DPMJET-III~\cite{dpmjet}. 
Earlier ATMNC code has been applied 
to the study of muon flux 
at several altitudes, at sea level, mountain altitude, and at balloon altitudes, where accurate measurements exist. The Monte Carlo generator with JAM shows a better agreement
than the former one without JAM.
Then it is applied to the calculation of atmospheric neutrino flux 
for several sites as the GranSasso, SNO, Kamioka and 
others~\cite{honda1,honda2,honda3}.

\begin{figure}
\includegraphics[width=14cm,height=10cm]{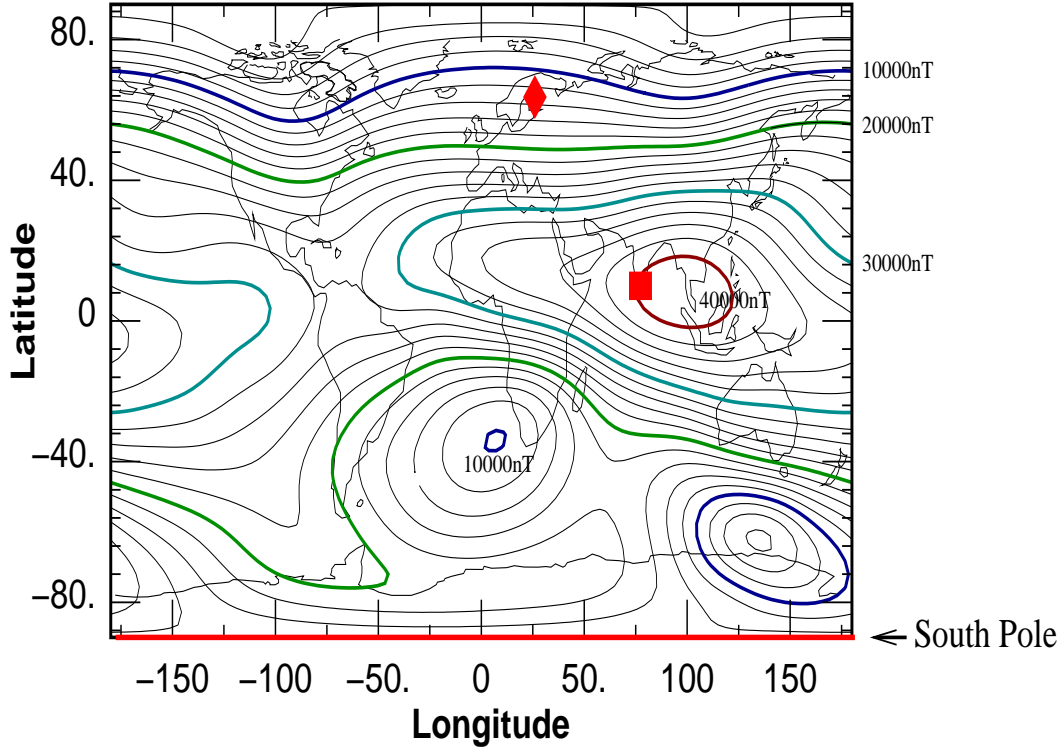}
\caption{Magnitude of the horizontal component of geomagnetic field in
IGRF2010 model~\cite{geofield}. Square stands for the position of India-based
Neutrino Observation (INO) site, diamond for the Pyh\"asalmi mines, and
bottom bar for the South Pole.}
\label{magnetic}
\end{figure}

In spite of using the same primary flux model and the interaction model 
for the different sites, 
the calculated atmospheric neutrino fluxes are different due to 
the geomagnetic field. 
The geomagnetic field affects cosmic rays both inside and outside of 
the atmosphere. 
First, it acts as a filter for low energy cosmic rays, 
and secondly, it deflects the charged particles in the atmosphere. 
These two effects are mainly controlled by the horizontal component 
of the geomagnetic field. 
It would be interesting to study the atmospheric neutrino flux 
for the three sites with different position in the geomagnetic field.
 In Fig.~\ref{magnetic}, we show the strength of the horizontal component of geomagnetic field obtained using IGRF2010 model~\cite{geofield} with the position of these sites.
 It can be seen that the INO site is close to the region where the strength of
the horizontal component of geomagnetic field is the largest on the earth.
The South Pole is close to the magnetic pole, where the horizontal 
component is zero.
At Pyh\"asalmi mine, the horizontal component of geomagnetic field is also small but not zero as it is little far away from the magnetic pole.

We proceed as follows. 
In section-\ref{intn-model}, we present 
the main features of the calculation scheme and in section-\ref{nuflx}, 
the results of the atmospheric neutrino fluxes have been shown and discussed. 
Finally, in section-\ref{concl}, we summarize the results and conclude 
our findings.

\section{CALCULATION SCHEME}\label{intn-model}
The scheme for calculating the atmospheric neutrino fluxes has been 
discussed in detail in the earlier work~\cite{honda1,honda2,honda3}.
We present here the main features. 
We use the primary flux model based on AMS~\cite{AMS1,AMS2} and 
BESS~\cite{BESS1,BESS2} data.
We use IGRF2010~\cite{geofield} model for the geomagnetic field. For the atmosphere model, we use NRLMSISE-00~\cite{atm-model} instead of US-standard76~\cite{nasa} which was used in our 
earlier works~\cite{honda1,honda2,honda3}.
 Actual calculation is carried out in the Cartesian coordinate system which
has the origin at the center of the Earth, with the Z-axis extending 
to the north pole, and we consider the surface of
the Earth to be a sphere with a radius of $R_e=6378.180$ km.
However, the position on the Earth is well represented
by the spherical polar coordinate system ($r, \theta, \phi$) 
with $r=R_e$ which is related to the Cartesian coordinate system by
\[x=R_e~\sin\theta~\cos\phi,~x=R_e~\sin\theta~\sin\phi ~{\text and}~ z=R_e~\cos\theta.\]

The local coordinate system at the detector is constructed
based on this polar coordinate system. The direction of the
x-axis is in the increasing direction of $\theta$, the direction of
the y-axis is in the increasing direction of $\phi$, and the
direction of the z-axis is in the increasing direction of r.
Therefore, the azimuth angle is measured counterclockwise
from south in the local coordinate system. 
In addition to the surface of the Earth, we assume three
more spheres; the injection sphere, the simulation sphere,
and the escape sphere. We have taken the radius of the injection sphere as 
$R_{\text inj}=R_e~+~100km$, and 
the radius of simulation sphere and the escape sphere are taken to be 
$R_{\text esc}=R_{\text sim}=10~\times~R_e$=63781.80 km.

Cosmic rays are sampled on the injection sphere uniformly towards 
the inward direction, following the given primary cosmic ray spectra. 
Before they are fed to the simulation code for the propagation in air, 
they are tested to determine whether they pass the rigidity cutoff or not. 
For a sampled cosmic ray, the ''history'' is examined by solving 
the equation of motion in the negative time direction. 
When the cosmic ray reaches the escape sphere without touching the injection
sphere again in the inverse direction of time, 
the cosmic ray can pass through the magnetic barrier following its trajectory
in the normal direction of time.
The propagation of cosmic rays is simulated in the space
between the surface of Earth and the simulation sphere.

We use the JAM interaction model for hadronic interactions below 32GeV, 
as this shows a better agreement with the HARP experiment~\cite{harp} 
and it agrees with the observed muon flux at sea level, 
at mountain altitudes and at balloon altitudes.
For energies above 32GeV, 
we use DPMJET-III~\cite{dpmjet} interaction model. 
We have checked the smooth interpolation when switching from 
the JAM model to the  DPMJET-III interaction model.
\begin{figure}
\includegraphics[width=14cm,height=10cm]{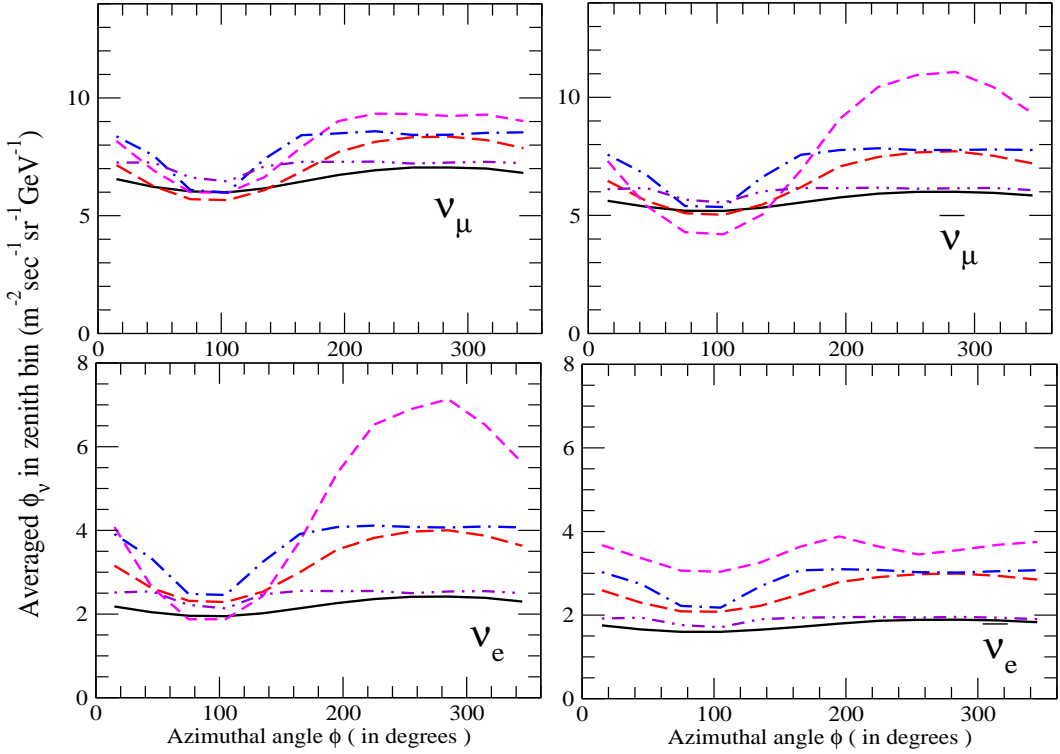}
\caption{The azimuthal angle dependence of atmospheric neutrino flux, 
averaged over zenith angle bins
of 1~$>~\cos\theta~>$~0.6 (solid line),~0.6~$>~\cos\theta~>$~0.2 (long dashed),
~0.2~$>~\cos\theta~>$-0.2 (short dashed),~ 
-0.2~$>~\cos\theta~>$~-0.6 (dashed-dotted),~and~-0.6~$>~\cos\theta~>$~-1 
(dashed double-dotted),~calculated for the 
INO site at (anti)neutrino energy E=3.2GeV.}
\label{ino-azim-3.2}
\end{figure}
\begin{figure}
\includegraphics[width=14cm,height=10cm]{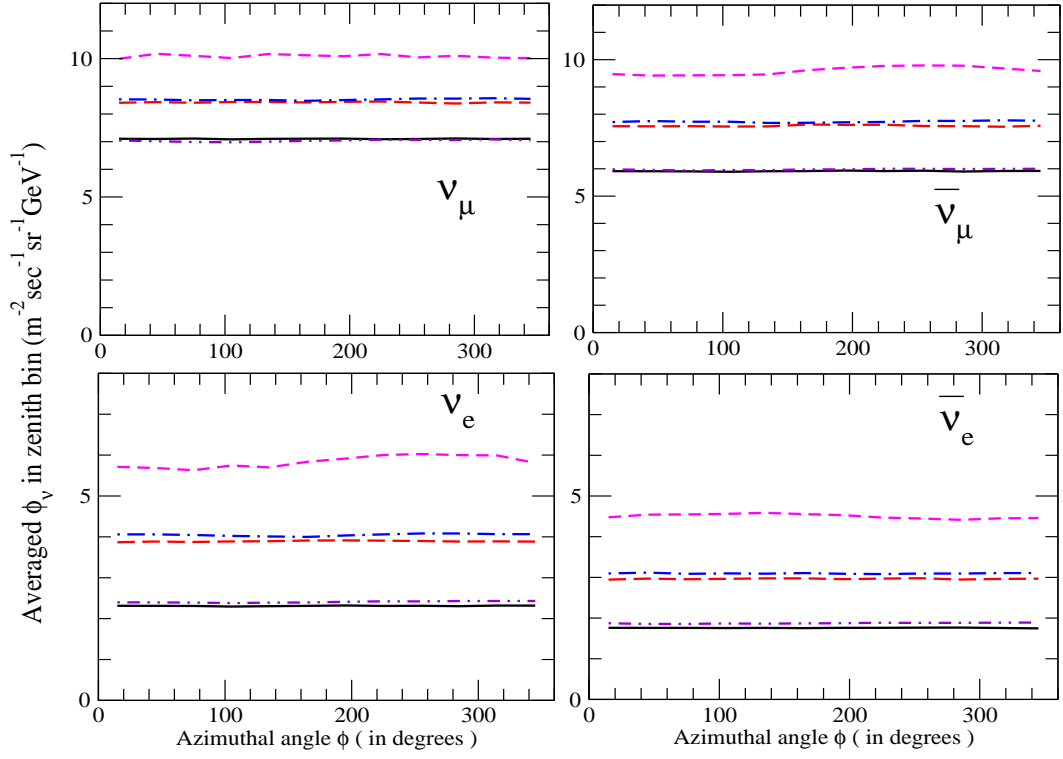}
\caption{Same as Fig.~\ref{ino-azim-3.2} at (anti)neutrino energy E=3.2GeV 
for the South Pole site.}
\label{sp-azim-3.2}
\end{figure}

\begin{figure}
\includegraphics[width=14cm,height=10cm]{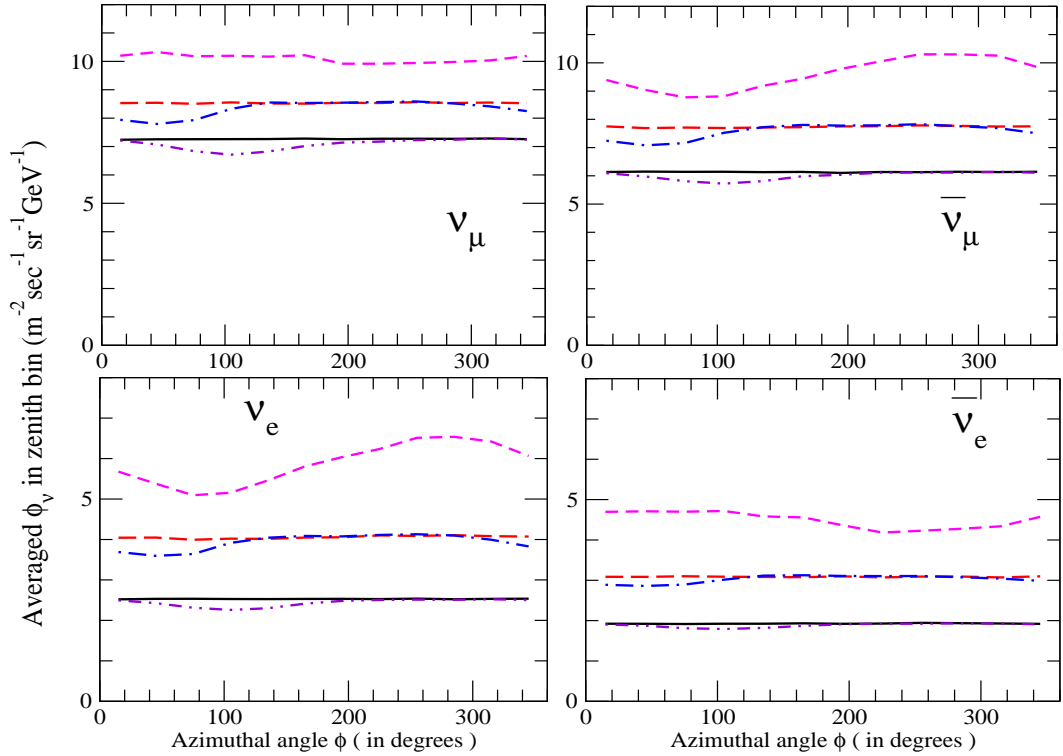}
\caption{Same as Fig.~\ref{ino-azim-3.2} at (anti)neutrino energy E=3.2GeV 
for the Pyh\"asalmi site.}
\label{py-azim-3.2}
\end{figure}

\begin{figure}
\includegraphics[width=14cm,height=10cm]{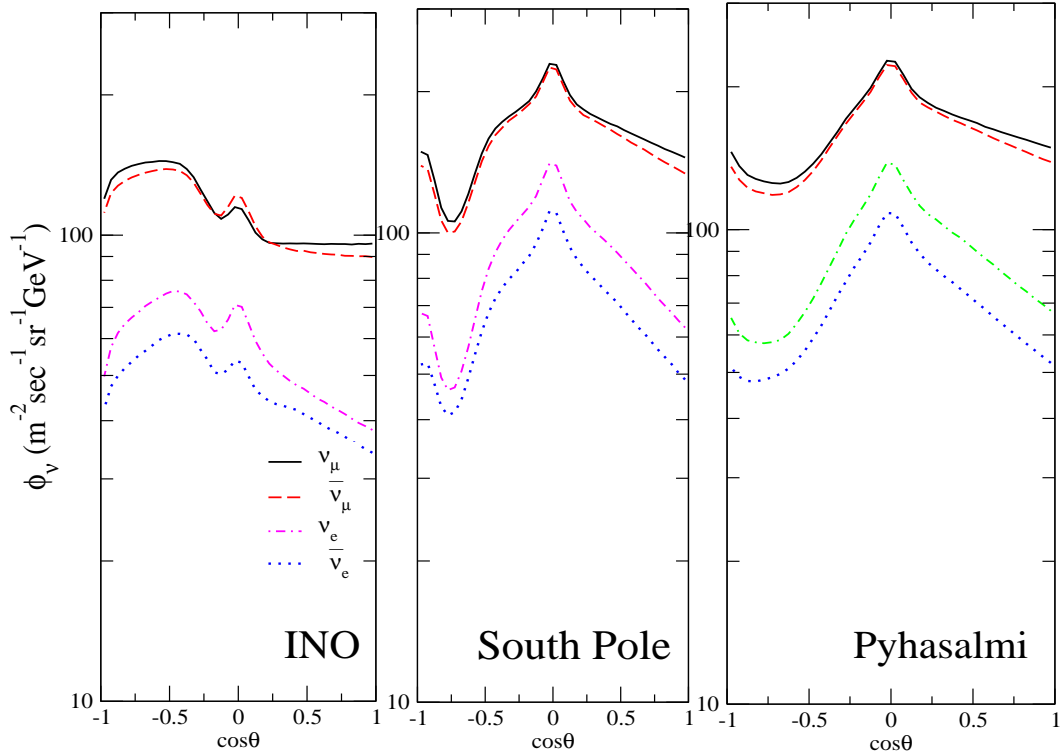}
\caption{The zenith angle dependence of atmospheric neutrino flux at E=1GeV, 
averaged over all azimuthal angles calculated for INO, South Pole 
and Pyh\"asalmi sites. 
Here $\theta$ is the arrival direction of the neutrino, with $\cos\theta=1$
for vertically downward going neutrinos, and $\cos\theta=-1$ for
vertically upward going neutrinos.}
\label{zenith-var-1.0}
\end{figure}

\begin{figure}
\includegraphics[width=14cm,height=10cm]{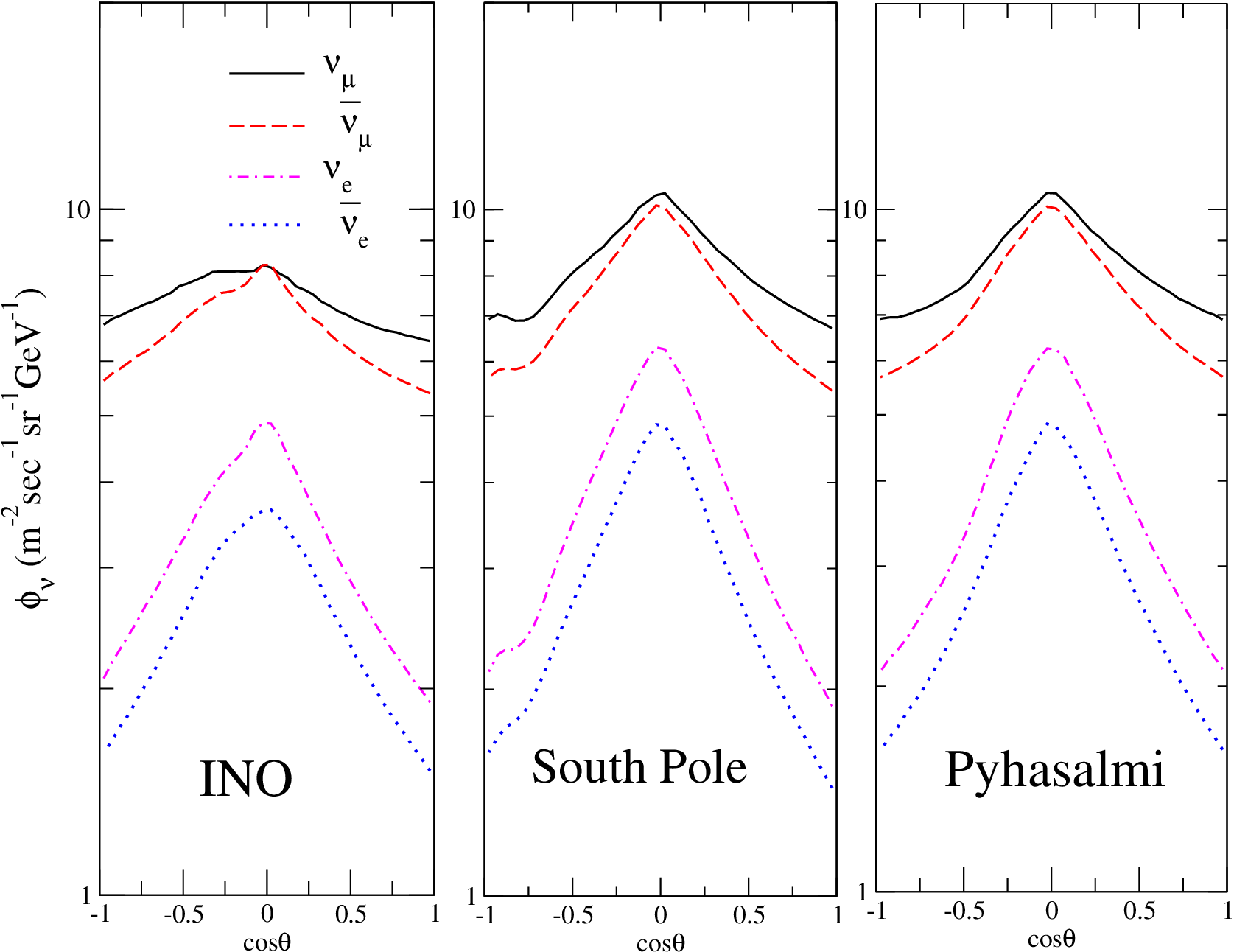}
\caption{Same as Fig.\ref{zenith-var-1.0} at E=3.2GeV.}
\label{zenith-var-3.2}
\end{figure}

\begin{figure}
\includegraphics[width=14cm,height=10cm]{zenith_var_10GeV.eps}
\caption{Same as Fig.\ref{zenith-var-1.0} at E=10GeV.}
\label{zenith-var-10}
\end{figure}

\begin{figure}
\centerline{
\includegraphics[width=7cm]{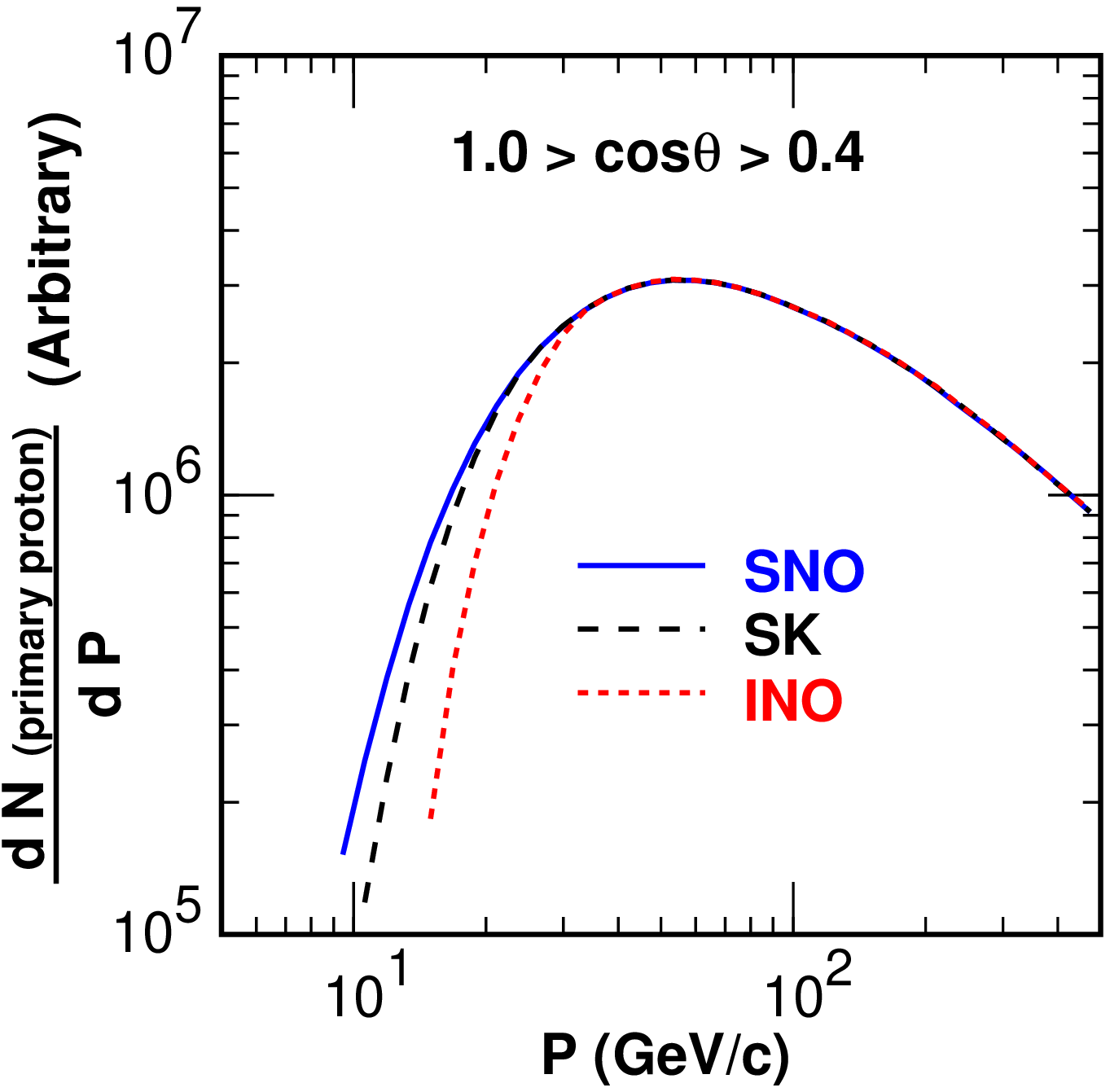}
\hspace{5mm}
\includegraphics[width=7cm]{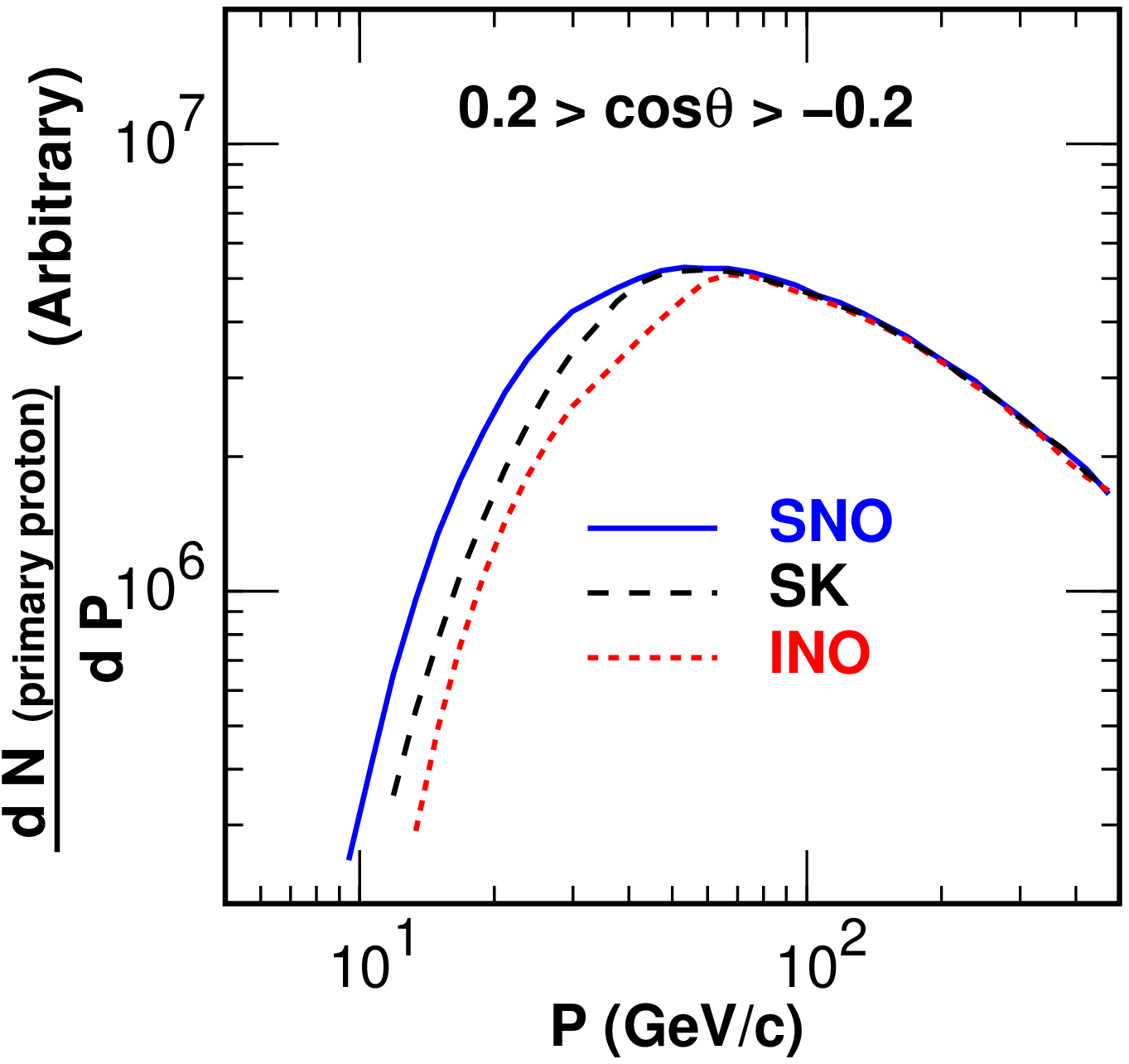}
}
\caption{The spectra of protons which create the neutrino above 3.2~GeV
from the near vertical ($ 1>\cos\theta_{zenith}>0.4 $, left panel)
 and near horizontal ($ 0.2>\cos\theta_{zenith}>-0.2$, right panel)
 directions at effectively no rigidity cutoff site (SNO~\cite{sno}), intermediate
rigidity cutoff site (SuperK~\cite{sk}) and at the INO site~\cite{ino}.}
\label{cutoff}
\end{figure}

\begin{figure}
\includegraphics[width=14cm,height=6cm]{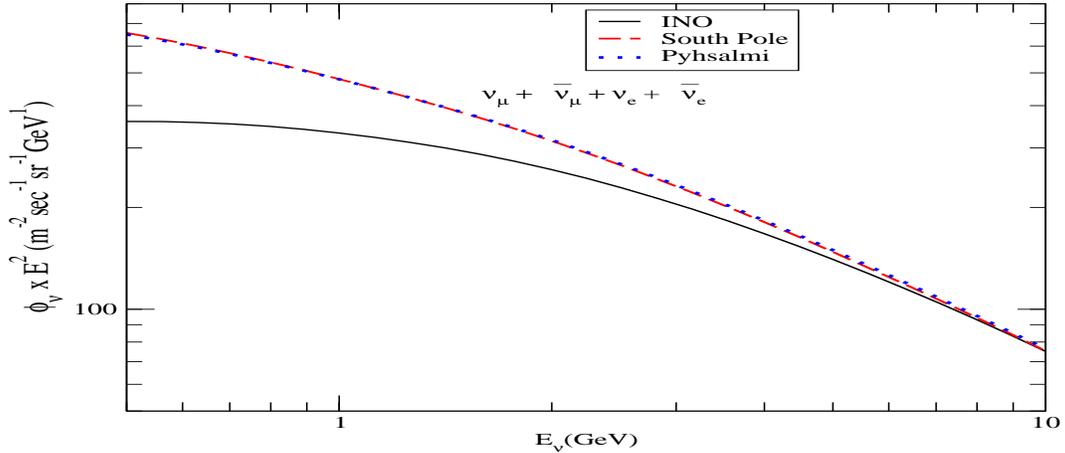}
\caption{Atmospheric neutrino flux sum over all zenith and 
azimuthal angles for the INO, South Pole and Pyh\"asalmi sites. 
These results are summed over all the flavors of neutrinos viz. 
$\nu_\mu~+~{\bar\nu}_\mu~+~\nu_e~+~{\bar\nu}_e$.}
\label{sumflx2}
\end{figure}

\section{Results and Discussions}
\label{nuflx}
In this section, we present the results of simulation for the 
atmospheric neutrinos at INO, South Pole and Pyh\"asalmi sites. 
 First we present the results for the atmospheric neutrino fluxes 
as a function of azimuthal angle $\phi$. 
These results are presented for 
$\nu_\mu$,~$\bar\nu_\mu$,~$\nu_e$~and~$\bar\nu_e$ for (anti)neutrinos of 3.2GeV.
In Fig.~\ref{ino-azim-3.2}, we present the results for the INO site, 
in Fig.\ref{sp-azim-3.2} the results are presented for the 
South Pole site and in Fig.~\ref{py-azim-3.2} the results are 
presented for the Pyh\"asalmi site at $E_\nu$=3.2GeV. 
In these figures we show the variation of atmospheric neutrino flux as the 
function of the azimuthal angle averaging them over the five zenith angle
ranges, 
1$~>~\cos\theta~>~$0.6, 
0.6$~>~\cos\theta~>$~0.2, 
0.2$~>~\cos\theta~>~$-0.2, 
-0.2$~>~\cos\theta~>~$-0.6 and 
-0.6$~>~\cos\theta~>~$-1. 
We find that the variation of the atmospheric neutrino flux has a 
complex structure at low (anti)neutrino energies, due to the 
rigidity cutoff and muon bending in the geomagnetic field. 
This variation remains almost the same for the near horizontal direction even above 10 GeV.
 Due to the high rigidity cutoff at the INO site 
this variation is more complex than the other two sites 
discussed here and for the South Pole site this variation is the least. 

Furthermore, the variation of upward going neutrinos is much more complicated
than the variation of downward going neutrinos. 
This is due to the fact that the upward-going neutrinos are produced in
wide area on the Earth, and there are large variation of rigidity cutoff 
and geomagnetic field.
 On the other hand, the downward going neutrinos are produced just above the
detector.

In Figs.~\ref{zenith-var-1.0},~\ref{zenith-var-3.2} and \ref{zenith-var-10}, 
we present the results for the atmospheric neutrino fluxes
as a function of the zenith angle after averaging over all the azimuthal
angles. 
The results are presented for neutrino energy of 1GeV, 
3.2GeV and 10GeV respectively, and for the three sites viz. 
INO~\cite{ino}, South Pole~\cite{pingu} and Pyh\"asalmi~\cite{pyh} 
in each of these figures. 
At 1 GeV, there are large up-down asymmetries in the atmospheric
neutrino flux at all the three sites. The downward going neutrino flux is
larger at the South pole~\cite{pingu} and Pyh\"asalmi~\cite{pyh} sites, while upward going neutrino flux is
larger at the INO site~\cite{ino} due to the different rigidity cutoff.
These asymmetries decrease with the increase in neutrino energy, and almost
disappear at 10 GeV. However, there appears an up-down asymmetry at South Pole,
due to the difference in the observation altitude.

We note that the horizontal/vertical flux ratio for the INO site is much 
smaller than for the other sites even at 3.2~GeV. 
This could be understood by the fact that the rigidity cutoff
still affect the neutrino flux at 3.2~GeV at the INO site,
and the rigidity cutoff is more effective in the horizontal direction.

In Fig.~\ref{cutoff}, we have shown proton spectra which produce 
 neutrino above 3.2 GeV, from the near vertical ($ 1>\cos\theta_{zenith}>0.4 $) and
near horizontal ($ 0.2>\cos\theta_{zenith}>-0.2 $) directions for the INO
site, with effectively no rigidity cutoff site (SNO~\cite{sno}) and intermediate
rigidity cutoff site (SK~\cite{sk}). On comparing the proton spectra for these
sites, it may be noticed that the rigidity cutoff works more efficiently
for the near horizontal direction than for the near vertical direction,
especially for the INO site~\cite{ino}. The rigidity cutoff for downward going neutrino at South pole~\cite{pingu}
and Pyh\'asalimi sites~\cite{pyh} are effectively the same as that of SNO site~\cite{sno}. 

In Fig.~\ref{sumflx2}, 
we present the results for the atmospheric neutrino spectra averaged 
over zenith and azimuth angles, for (anti)neutrino energies 
from 0.5 GeV to 10 GeV, for the INO~\cite{ino}, South Pole~\cite{pingu} 
and Pyh\"asalmi~\cite{pyh} sites. 
We find that when the neutrino flux is integrated over all the angles, 
the difference in the flux at the South Pole and Pyh\"asalmi sites 
which appeared in Figs.~\ref{zenith-var-1.0},~\ref{zenith-var-3.2} 
and \ref{zenith-var-10}, disappears. Due to the strong effect of the horizontal
 component of the geomagnetic field atmospheric neutrino flux 
at the INO site is almost 30$\%$ smaller at 1GeV, than the flux at the 
two other sites discussed here, 
while this difference becomes smaller with the increase in the neutrino energy, 
for example, at 3GeV, this difference is 10$\%$, 
but the flux reduces almost by a factor of 20 as compared to the flux at 1GeV.
\section{SUMMARY AND CONCLUSION}\label{concl}
We have studied the atmospheric neutrino flux for the INO, 
South Pole and Pyh\"asalmi sites using 
ATMNC with JAM interaction code below 32 GeV. 
We find that 
the atmospheric neutrino flux is quite different in nature particularly 
at low and intermediate energies,
depending on the position in the geomagnetic field, 
and the strength of the horizontal component of the geomagnetic field is a good measure of the deviation in the fluxes at the different sites.
Especially, the difference of the zenith angle dependence of 
the atmospheric neutrino flux is important in the analysis of 
the neutrino oscillation.
The difference is large at lower energies
up to the neutrino energies of a few GeV.
These results would be useful in the analysis of the atmospheric neutrino 
experiments proposed at these sites.
\begin{acknowledgments}
One of us(MSA) would like to thank the University of Tokyo for the financial support to visit and work at ICRR.  
Computer facility at the ICRR is greatly acknowledged.
\end{acknowledgments}

\end{document}